\documentclass[useAMS,usenatbib]{mn2e} 
\usepackage{aas_macros}
\usepackage{graphics}
\usepackage[pdftex]{graphicx}
\usepackage{epstopdf}
\usepackage{epsfig}  
\usepackage{natbib} 
\usepackage{dingbat}
\usepackage{float}
\usepackage{amsmath}
\usepackage{times}
\usepackage[varg]{txfonts}
\usepackage{verbatim} 
\usepackage{multirow,bigdelim} 
\usepackage{color}
\usepackage{vmargin}
\setmarginsrb{1.2cm}{1cm}{1.2cm}{1cm}{1cm}{1cm}{1cm}{1cm}

  \makeatletter
    \renewcommand{\paragraph}{\@startsection{paragraph}{4}{\z@}%
      {-3.25ex\@plus -1ex \@minus -.2ex}%
      {1.5ex \@plus .2ex}%
      {\normalfont\small\centering}}
     
    \renewcommand{\subparagraph}{\@startsection{subparagraph}{5}{\z@}%
      {-3.25ex\@plus -1ex \@minus -.2ex}%
      {1.5ex \@plus .2ex}%
      {\normalfont\small\centering}}
    \makeatother

\newcommand{\kms}{{ km~s$^{-1}$}}
\newcommand{\hMpc}{{ h$^{-1}$~Mpc}}

\title[WF and Datasets]{Toward an Optimal Sampling of Peculiar Velocity Surveys For Wiener Filter Reconstructions}
\author[Sorce et al.]
{Jenny G. Sorce$^{1,2}$\thanks{E-mail: \text{jsorce@aip.de / jenny.sorce@astro.unistra.fr}}, 
Yehuda Hoffman$^{3}$,
Stefan Gottl\"{o}ber$^1$\\
$^1$Leibniz-Institut f\"{u}r Astrophysik, 14482 Potsdam, Germany\\
$^2$Universit\'e de Strasbourg, CNRS, Observatoire astronomique de Strasbourg, UMR 7550, F-67000 Strasbourg, France\\
$^3$Racah Institute of Physics, Hebrew University, 91904 Jerusalem, Israel\\
}
\begin{document}

\date{}

\pagerange{\pageref{firstpage}--\pageref{lastpage}} \pubyear{2017}

\maketitle

\label{firstpage}

\begin{abstract}
\indent The Wiener Filter (WF) technique enables the reconstruction of density and velocity fields from observed radial peculiar velocities. 
This paper aims at identifying the optimal design of peculiar velocity surveys within the WF framework. The prime goal is to test the dependence of the quality of the reconstruction on the distribution and nature of data points. Mock datasets, extending to 250~\hMpc, are drawn from a constrained simulation that mimics the local Universe to produce realistic mock catalogs. Reconstructed fields obtained with these mocks are compared to the reference simulation. Comparisons, including residual distributions, cell-to-cell and bulk velocities, imply that the presence of field data points is essential to properly measure the flows. The fields reconstructed from mocks that consist only of galaxy cluster data points exhibit poor quality bulk velocities. In addition, the quality of the reconstruction depends strongly on the grouping of individual data points into single points to suppress virial motions in high density regions. Conversely, the presence of a Zone of Avoidance hardly affects the reconstruction. For a given number of data points, a uniform sample does not score any better than a sample with decreasing number of data points with the distance. The best reconstructions are obtained with a grouped survey containing field galaxies: Assuming no error, they differ from the simulated field by less than 100~\kms\ up to the extreme edge of the catalogs or up to a distance of three times the mean distance of data points for non-uniform catalogs. The overall conclusions hold when errors are added.

\end{abstract}


\section{Introduction}

Reconstructing the local Large Scale Structure is essential in order to analyze the distribution of matter and to understand the motions ruling the local Universe, namely to study the local dynamics. Several methods and algorithms have been devised for that purpose over the last three decades \citep[e.g.][]{1990ApJ...364..349D,1990ApJ...364..370B,1992ApJ...391..443N,2008PhyD..237.2139L,2013MNRAS.429L..84K,2013MNRAS.432..894J,2013ApJ...772...63W,2016MNRAS.457..172L}. Leading efforts in that direction are the POTENT analysis \citep{1999ApJ...522....1D} and the Wiener Filter (WF) Bayesian methodology \citep{1995ApJ...449..446Z,1999ApJ...520..413Z}. These techniques can usually be applied either to radial peculiar velocity catalogs or to redshift surveys. While the latter are easily acquired they provide a biased account of the matter distribution. On the other hand, radial peculiar velocity acquisition constitutes a real observational challenge, but unlike redshift surveys, they are a better tracer of the total (including the dark component) mass distribution. In addition, peculiar velocities are linear on scales of a few megaparsecs and are correlated over large distances. It follows that the reconstruction of the Large Scale Structure from radial velocities constitutes a challenging and appealing task to observers and theorists alike. 

Determination of galaxy peculiar velocities is uncertain primarily due to the distance measurement (i.e. the Hubble flow subtraction). Extragalactic distance measurement is exceptionally difficult and telescope time consuming. Hence, strategically, the optimal survey must be defined to plan the observations accordingly. In this paper, we focus on the particular aspect of data sampling within the framework of the WF reconstruction. We address questions such as: Should field galaxies be the subject of all the efforts or should galaxy clusters be the focus? What are the merits, if any, of using clusters as individual data points? To what extent does the Zone of Avoidance (ZOA) degrade the quality of the reconstructions \citep{1994ASPC...67..185H}? Should one opt for a uniform spatial sampling?. 

The modus operandi of this paper is to build mock catalogs out of a constrained simulation of the local Universe \citep[e.g.][]{2010arXiv1005.2687G,2014MNRAS.437.3586S}. The advantage of using a constrained simulation is that the simulation reproduces all the known major structures and therefore realistic mock datasets can be constructed. Dark matter halos and sub-halos are used as proxies for clusters and galaxies. To dedicate our attention entirely to the sampling issues, neither statistical nor systematic errors are added to distances and peculiar velocities of halos in a first-pass. In order to build the mock catalogs, given the constrained nature of the simulation, the observer is assumed to stand in the center of the box where the origin of the Supergalactic coordinate system is positioned. The second peculiar velocity catalog  \citep[{\it cosmicflows-2},][]{2013AJ....146...86T}  of the cosmicflows project is used here as our benchmark for peculiar velocity surveys. It has actually already been used in a series of paper to study the local Large Scale Structure, to set up initial conditions for constrained simulations and to estimate the bulk flow \citep[e.g.][]{2014Natur.513...71T,2015MNRAS.449.4494H,2015MNRAS.450.2644S,2015ApJ...812...17P,2015MNRAS.447..132W,2016MNRAS.455.2078S,2016MNRAS.458..900C,2016MNRAS.460L...5C,2016MNRAS.460.2015S,2016MNRAS.461.4176H}. 

In its modern form the WF is formulated as a Bayesian linear estimator based on derivation of correlation functions, matrices and their inverse assuming a prior cosmological model. It is optimal in the case of Gaussian random fields \citep[][and see Appendix A for detailed equations]{1995ApJ...449..446Z}. The linear regime of the Large Scale Structure is one of these explaining why the WF has revealed itself to be such a vital tool for reconstructing the Large Scale Structure. It is then natural to study the optimal sampling strategy within the WF framework.  The WF is a useful estimator in the case of very noisy, sparse, unhomogeneously distributed with an incomplete coverage datasets, namely with catalogs like peculiar velocity surveys. It can handle all the sampling issues previously described and this paper aims at quantifying to which level it does so. However, there is one more issue that needs to be tackled with: the grouping or collapse of all the data points that belong to one given large halo (cluster) into one point.  This procedure is used to suppress non-linear virial motions that are not accounted for in the WF.  This paper tests the importance of removing non-linear motions.

The paper is divided as follows. In the second section different mock catalogs that can be distinguished by 1) the number of data points as a function of distance, 2) the presence or not of a ZOA, 3) the selection of clusters only, 4) mimicking a grouping scheme while keeping isolated field galaxies, are drawn from the simulation. In the third section, the WF technique is applied to these different mocks and the reconstructed velocity fields are compared to the velocity field of the reference simulation. In particular, reconstructed dipole and monopole terms are compared to those of the simulation. Before concluding, a brief excursion in the regime of radial peculiar velocity catalogs with uncertainties is undertaken.

\begin{figure*}
\vspace{-0.5cm}
\includegraphics[width=0.8 \textwidth]{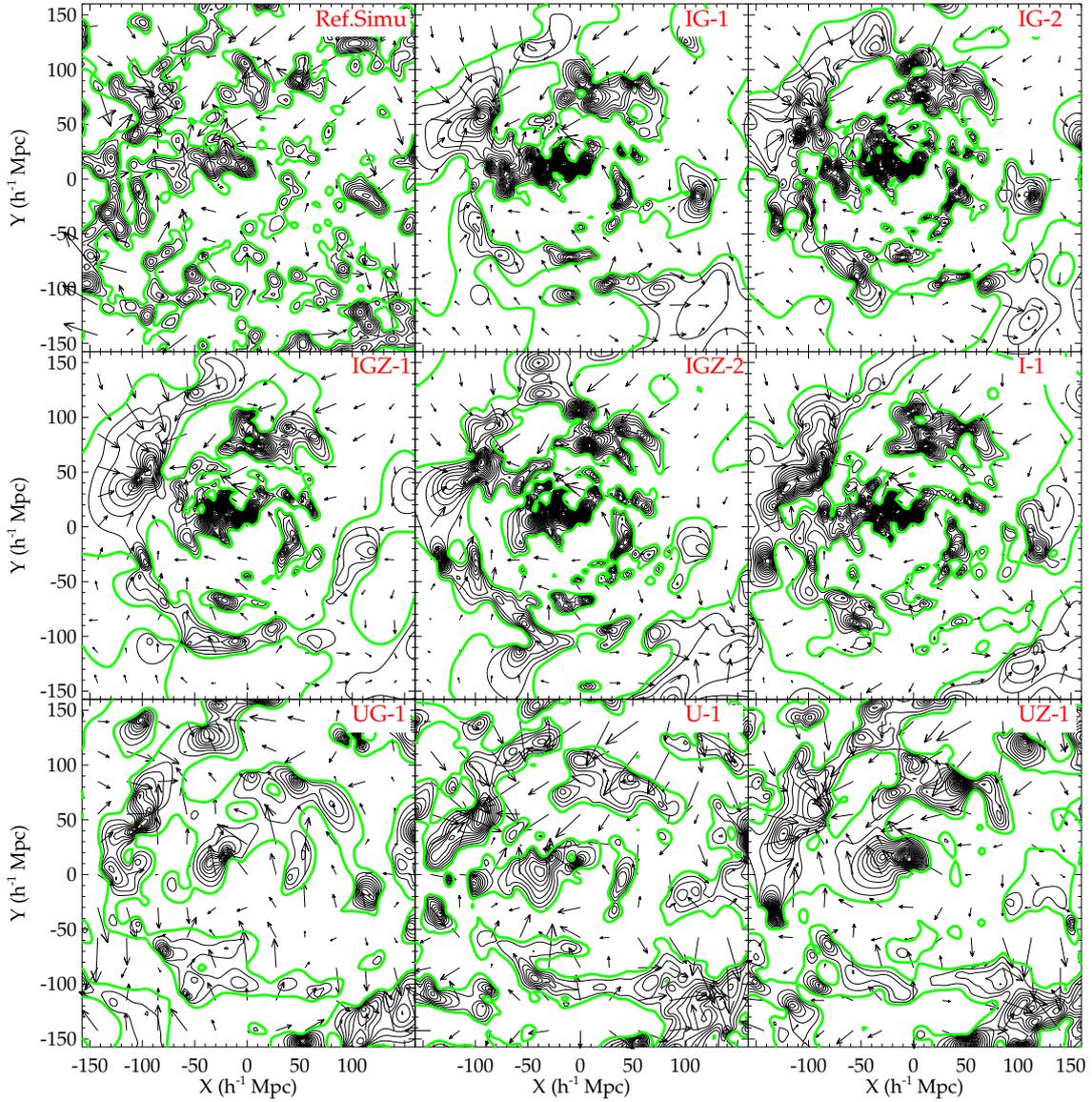}
\caption{First top, left panel: Density (black, green color for the mean density) and velocity (black arrows) fields of the reference simulation in the XY supergalactic plane. Other panels: Reconstructed overdensity (black and green contours) and velocity (black arrows) fields obtained with the Wiener Filter technique applied to one mock of each category. The properties of these mocks and the corresponding initials are explained in Table \ref{Tbl:1} in the same order. The Large Scale Structure of the reference simulation is overall reconstructed in every case.}
\label{fig:wf}
\end{figure*}


\section{Building Mock Catalogs}
\subsection{The reference simulation}

To be able to meet the requirements of the benchmark catalog in the tests, a constrained simulation of the local Universe is made in the context of the CLUES project\footnote{https://www.clues-project.org/} \citep{2010arXiv1005.2687G} following the process described in \citet{2014MNRAS.437.3586S,2015MNRAS.450.2644S}. This simulation is 500~\hMpc\ wide and contains 1024$^3$ particles. The Planck cosmology framework \citep[$\Omega_m$=0.307, $\Omega_\Lambda$=0.693, H$_0$=67.77, $\sigma_8=0.829$,][]{2014A&A...571A..16P}, used to run the simulation, serves as the prior model in the WF.

Unlike the more typical cosmological simulations that derive from random realizations of the matter power spectrum, constrained simulations stem from a set of constraints that can be either redshift surveys or radial peculiar velocities. The simulation used here obey a set of radial peculiar velocities. Most importantly, a cloud-in-cell scheme revealed that a look alike for all the major structures and voids of the local Universe can be found in this simulation ensuring that the large scale environment is similar to our neighborhood \citep{2016MNRAS.455.2078S}. In Figure \ref{fig:wf}, we consider the box to be equivalent to the local Universe, an observer is placed at the center of the box, coordinates can be defined similarly to observational supergalactic coordinates. In this set of coordinates, the XY plane is shown in the top left panel of Figure \ref{fig:wf}. The velocity (black arrow) and density (black contour, green color for the mean) fields of the reference simulation are represented. In that simulation all the major attractors and voids of the local Universe are represented: the Shapley supercluster in the top left, the Coma supercluster in the middle top, the Virgo supercluster in the middle of the box and the Perseus Pisces supercluster on the right side of Virgo.

\subsection{Building Mocks}

\begin{table*}
\begin{center} 
\begin{tabular}{cccccccc}
\hline
\hline
Category & Uniform distribution & Inhomogeneous distribution & ZOA & Ungrouped & Grouped+field galaxies & Clusters\\
\hline
IG-X & & \checkmark & &  & \checkmark &\\
IGZ-X & & \checkmark &  \checkmark &&\checkmark & \\
I-X & &\checkmark & &   \checkmark & &\\
UG-X & \checkmark & & & & &\checkmark  \\
U-X & \checkmark & & &  \checkmark &&  \\
UZ-X & \checkmark & & \checkmark &\checkmark && \\
\hline
\hline
\end{tabular}
\end{center}
\vspace{-0.25cm}
\caption{Different mock types: 'U' stands for a uniform distribution, these mocks match {\it cosmicflows-2} (either grouped 'G' or ungrouped although a uniform distribution gives rise to an almost grouped catalog) solely by the number of data points. Note that only clusters are selected in the case of the UG mocks while in the case of U mocks, the selection is completely random. 'I' stands for inhomogeneous distribution, namely not only the number of data points is mimicked but also their distribution as a function of the distance. 'Z' stands for the presence of a zone devoid of data mimicking the Zone of Avoidance. Finally 'X' can take the value 1 or 2 depending on whether there is once or twice the number of data points in {\it cosmicflows-2} grouped and ungrouped respectively.  For each category 5 mocks that differ on the 'randomly' chosen halos are built. Note however that, since there are approximately as many halos as number of measurements in the center of the box because of the simulation resolution, the number of variations possible is smaller when mimicking also the distribution of halos in {\it cosmicflows-2} especially when doubling the number of measurements.}
\label{Tbl:1}
\end{table*}

\begin{figure*}
\vspace{-0.5cm}
\includegraphics[width=0.8 \textwidth]{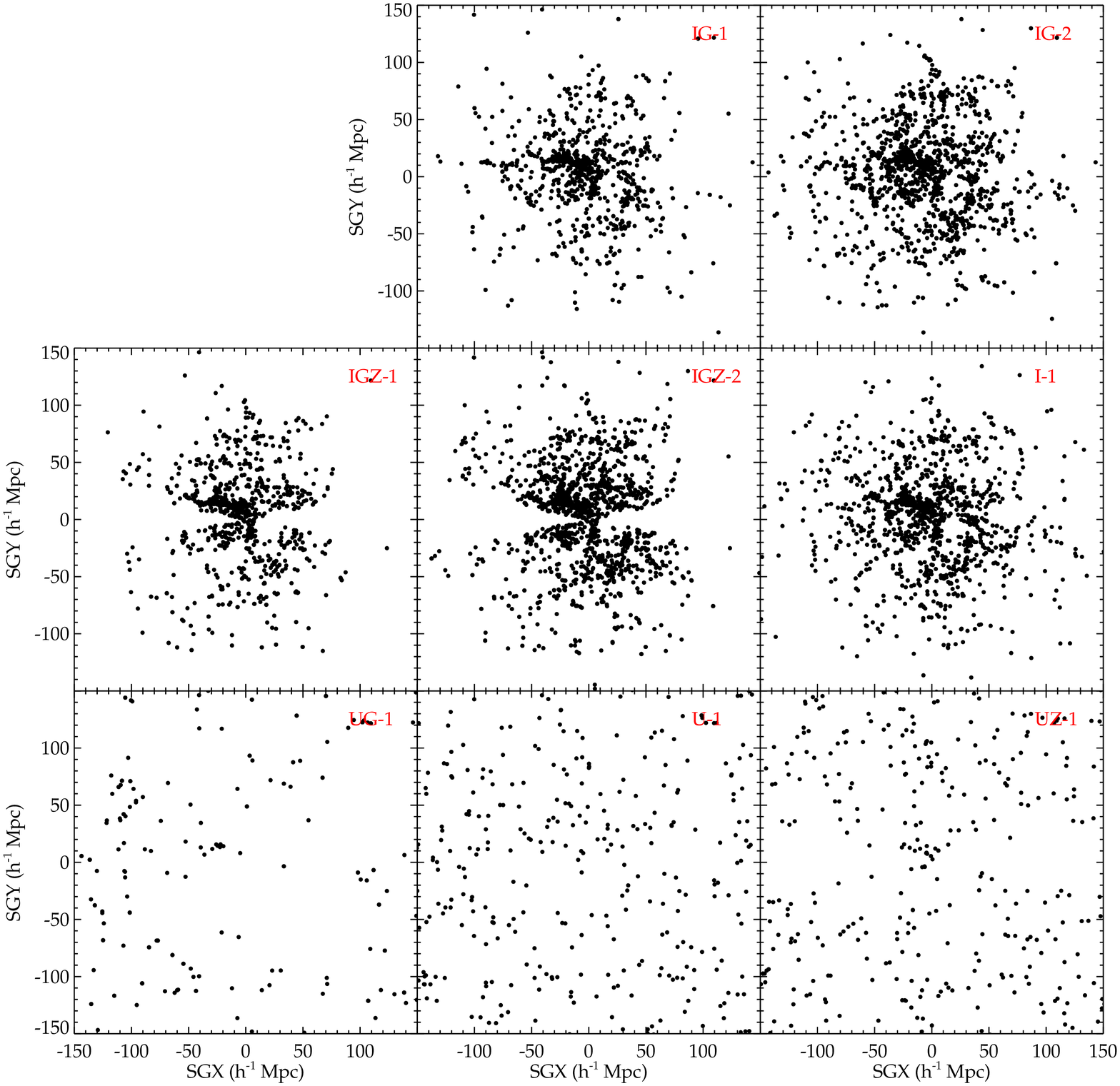}
\caption{Distributions of selected halos in the XY supergalactic plane in a 10~\hMpc\ thick slice of one mock per each category. From left to right, top to bottom: distribution similar to the grouped version of {\it cosmicflows-2} without a Zone of Avoidance (ZOA) and with the same number of data points (IG-1), with twice the number of data points and still without ZOA (IG-2), with the same number of data points and a ZOA (IGZ-1), with twice the number of data points and with a ZOA (IGZ-2); repartition matching that in ungrouped {\it cosmicflows-2} without ZOA (I-1); uniform distribution matching only the number of points in the grouped version of {\it cosmicflows-2} and selecting only massive halos (UG-1), uniform distribution matching the number of points in ungrouped {\it cosmicflows-2} without  (U-1) and with a ZOA (UZ-1). }
\label{fig:mocks}
\end{figure*}

The second generation catalog of galaxy distances built by the Cosmicflows collaboration\footnote{http://www.ipnl.in2p3.fr/projet/cosmicflows/} is a large publicly released catalog of radial peculiar velocities that will be considered as the benchmark to build the various mocks. Published in \citet{2013AJ....146...86T}, it contains more than 8,000 accurate galaxy peculiar velocities. Distance measurements come mostly from the Tully-Fisher \citep{1977A&A....54..661T} and the Fundamental Plane \citep{2001MNRAS.321..277C} methods. Cepheids \citep{2001ApJ...553...47F}, Tip of the Red Giant Branch \citep{1993ApJ...417..553L}, Surface Brightness Fluctuation \citep{2001ApJ...546..681T}, supernovae of type Ia \citep{2007ApJ...659..122J} and other miscellaneous methods also contribute to this large dataset though to a minor extent ($\sim 12\%$ of the data). Our primary interest lies in the total number of points and the repartition of these points as we consider the ideal case of perfect mock catalogs without any error, and as a result without any bias. The grouped version of {\it cosmicflows-2} is useful for the grouped mocks. It contains 552 groups and 4303 single galaxies, i.e a total radial peculiar velocity count of 4855.

We use the Amiga halo finder \citep[][]{2009ApJS..182..608K} to build a list of halos from our reference simulation. Halos are then selected and prepared to match {\it cosmicflows-2} on different aspects:
\begin{itemize}
\item Mocks with a uniform distribution only reproduce the total number of data points in {\it cosmicflows-2} and extend to the same maximum distance as {\it cosmicflows-2} (approximately 250~\hMpc). Note that the uniformity of the distribution gives rise to mocks that are de facto already grouped in the sense that there is not much to group when data points are uniformly distributed and sparse.
\item For mocks with an inhomogeneous distribution we seek to have a similar repartition of data points (number and spatial coverage) as {\it cosmicflows-2}. An histogram with a bin size of 20~\hMpc\ is derived for the observational {\it cosmicflows-2} catalog providing the number of measurements in each bin or 20~\hMpc\ shell. For each 20~\hMpc\ shell, the same number of halos as found in {\it cosmicflows-2} is selected randomly in the list of halos. 
\item For grouped mocks, in the case of multiple halos in the same region, the most massive halo is selected to mimic the grouping applied to {\it cosmicflows-2}. This ensures that 1) the velocity of the halo is free of non-linear motions that could be induced by more massive nearby halos and 2) the center of mass of the region corresponds approximately to the center of mass of the halo. Subsequently, its position and velocity are directly given by the halo finder. This procedure permits mimicking a group catalog without an extra layer of complexity based on the selection of a grouping algorithm and all its implications, namely defining the galaxies belonging to the group and the velocity and position to be attributed to the group. Note that for the grouped mock obtained with a uniform distribution, only the most massive halos (by extension clusters) are selected. This is only a subtle difference but with non negligible effects on the results.
\item To reproduce the ZOA, domains are identified as similar to the ZOA and every halo is removed from those. This zone is defined as a cone with the apex at the center of the box (where the observer is assumed to lie), with an aperture angle of 20$^\circ$ and assuming the same orientation within the XYZ volume as the observational one in the supergalactic XYZ volume. 
\end{itemize}
\vspace{-0.25cm}
The different types of mocks are summarized in Table \ref{Tbl:1}.\\

We produce five mocks of each type,i.e. in a category, mocks differ from one another only by the `randomly' chosen data points. In the case of the inhomogeneous distribution the variance is limited close to the center of the box since there are almost as many halos to select as halos available despite the resolution of the simulation. We also double the number of data points\footnote{Note that the variance in the center of the box is reduced some more.} to see the influence of having more observations in the grouped mocks mimicking the grouped version of {\it cosmicflows-2}, in terms of distribution of data points. This permits us to determine whether there is an advantage in having additionnal points. Since the perfect case without error is considered, there will be no quantitative measurements on the fact that having more points has the unconditional advantage of decreasing the errors. On the other hand, a group with no error measurements on velocities and positions can already be considered as the result of the grouping of an infinite number of measurements of the group-members.

On Figure \ref{fig:mocks}, we compile the resulting list of halos for each mock type (one mock per category is represented) is plotted as black dots in a 10~\hMpc\ thick slice in the XY supergalactic plane.

\section{Wiener Filter Reconstructions}

\subsection{Velocity fields in general}

\begin{figure*}
\vspace{-3cm}
\includegraphics[width=0.9 \textwidth]{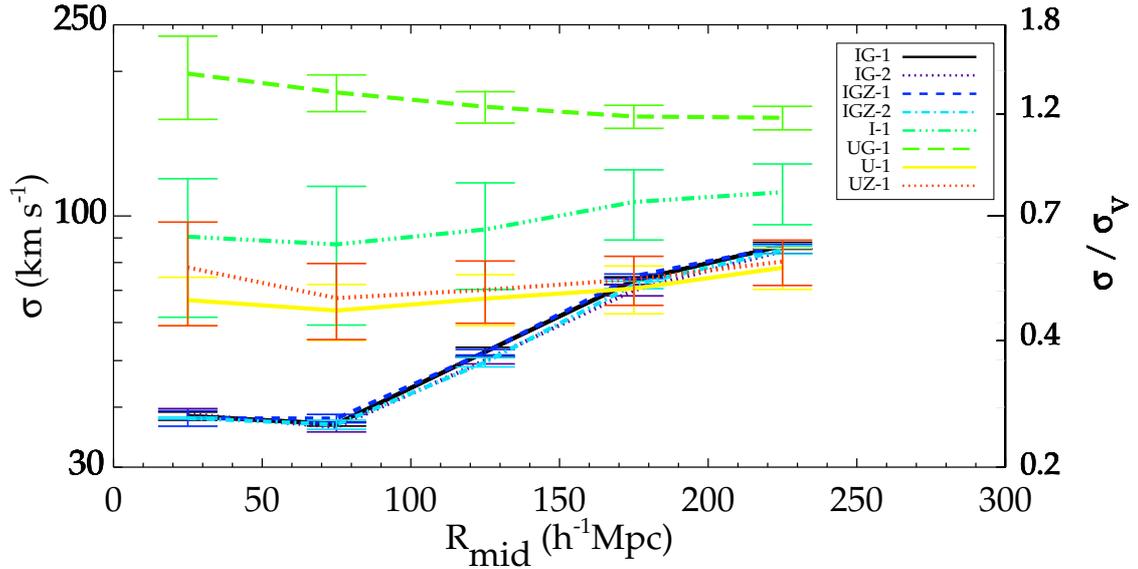}
\vspace{-2cm}

\caption{Means (lines) and standard deviations (error bars) of 1-$\sigma$ scatters obtained comparing the velocity field of the reference simulation and that obtained with the WF technique applied to the different mocks as a function of the `middle' radius of the compared shells, i.e. grid cells at a distance d such that R$_{mid}$-25$\le$~d~$<$~R$_{mid}$+25~\hMpc\ are compared. Each colored linestyle corresponds to a mock type where the short names refer back to Table \ref{Tbl:1}.}
\label{fig:scattersimu}
\end{figure*}

We apply the WF technique to the different mocks using the Planck power spectrum as a prior. The main calculations of the method are given in Appendix A. A boxsize of 500~\hMpc\ and grids with 256$^3$ cells are used. The resulting overdensity and velocity fields are plotted in the XY supergalactic plane in Figure \ref{fig:wf} with one mock of each category together with those of the reference simulation. Overdensity fields are shown as black contours (the green color represents the null field-contour) while the black arrows represent the velocity fields. Some differences can be observed between the different WF reconstructions although overall they all reproduce the Large Scale Structure of the reference simulation confirming that the WF is a powerful technique.

To assess to what extent, a reconstruction is in agreement with the reference simulation, we proceed as follows:
\begin{itemize}
\item production of a cell-to-cell comparison between the velocity field of the reference simulation and that of the reconstruction. Namely all the cell values from field 1 are plotted against those of field 2. Using a cloud-in-cell scheme with a 256$^3$ grid for the simulation to match the grid size of the WF reconstructions, the cell sizes are 1.95~\hMpc\ i.e. slightly smaller than the linear threshold.
\item derivation of the one sigma (1-$\sigma$) scatter of this cell-to-cell comparison, i.e. if the fields are identical, the points (cell field 1 ; cell field 2)  all lie on the 1:1 relation.
\item repeating of the process with the four reconstructions obtained with the other mocks in the same category as the first one,
\item computation of the mean and standard deviation of the 1-$\sigma$ scatters. 
\end{itemize}
Not to bias results towards a better agreement with mocks having more data points in the center, the cell-to-cell comparisons are made within shells of 50~\hMpc\ thickness (thus the center of the box is not always accounted for). The derived mean 1-$\sigma$ scatters are plotted together with their standard deviation (bars) on Figure \ref{fig:scattersimu} as a function of the `middle' radius R$_{mid}$ such that compared shells are constituted of cells at distances between R$_{mid}$-25 and R$_{mid}$+25~\hMpc\ from the center of the box. Each color represents one type of mocks as given by the legend and the meaning of the short names are described in Table \ref{Tbl:1}.

The first observation is that the standard deviations of the 1-$\sigma$ scatters are quite small from one reconstruction to another whatever shell is considered: from 0.2 to 39~\kms. The majority is around 1-2~\kms\ for the non-uniformly distributed mocks, with the exception of the ungrouped mocks that have standard deviations between 16 and 29~\kms. The non (but almost) grouped mocks with the uniform distribution present standard deviations of 8-10~\kms. The catalogs of clusters result in the maximum standard deviations with values between 9 and 39~\kms. Standard deviations are the smallest for mocks mimicking the grouped version of {\it cosmicflows-2} not only in terms of number of points but also in their repartition. That on average in all the shells, the highest standard deviations are those observed for the mocks mimicking an ungrouped configuration confirms the influence of non-linear motions on a linear technique such as the WF. More precisely, from one selection to another the non-linear motions are more or less important implying that they do not affect the reconstruction to the same amount. 

As a second observation, one can note that the presence of a ZOA does not impact the reconstruction severely, if it does at all, as already shown by \citet{1994ASPC...67..185H}: the 1-$\sigma$ scatters are about the same when comparing mocks with and without ZOA. They are about 85~$\pm$~1~\kms\ for the largest shells and decrease, with the size of the shells, down to 35-40~$\pm$~1~\kms\ when comparing reconstructions from inhomogeneous mocks. For uniform mocks, the 1-$\sigma$ scatters are more stable across the different shells with values about 80~$\pm$~8~\kms, in agreement with the fact that they are homogeneously distributed. When comparing the smallest shells, the effect of the non-linear clustering captured only by the simulation is visible on small scales and explains that 1-$\sigma$ scatters increase when comparing small volumes. This is the limitation of the method used here to compare the simulation with non linear motions to the linear WF reconstructions. Still, this method is efficient enough to reach our goal.

A slight but negligible ($<$~2~\kms) improvement is observed when using twice the number of points as that in {\it cosmicflows-2} only in the outer shells. This implies that increasing the density of data points improves the reconstruction only weakly comforting the stability of the WF technique. For the ungrouped and inhomogeneous catalogs, non-linear motions scramble the signal so that reconstructions only have accuracy at about 100~$\pm$~20~\kms\ minimum. 

With regards to the catalogs of clusters, they present the largest 1-$\sigma$ scatters of all the tested mock catalogs in agreement with expectations: they do not contain galaxies in the field to measure the gravitational field due to the attraction of large attractors like clusters. Such catalogs permit to recover the velocity field only at an accuracy of about 160~$\pm$~10~\kms\ for the largest shells up to 200~$\pm$~39~\kms\ for the inner shells. 

As a final note, the uniformly distributed catalogs (except those of clusters) result on average in slightly better reconstructions than the inhomogeneous grouped catalogs only for the largest shells. Still, because the variance of their 1-$\sigma$ scatters is larger, it includes the mean 1-$\sigma$ scatters and the variances obtained for the reconstructions made with the non-uniformly distributed grouped mocks. Namely, with less than 2\% of the data in the largest compared shells (at R$_{mid}$=225~\hMpc), the catalogs reproducing the distribution of the second catalog of the Cosmicflows project are able to do as well as the uniformly distributed catalogs that have about 30\% of data in that region.\\

To give another quantitative aspect of the mean 1-$\sigma$ scatters, their ratio to the variance of the simulated field is added as a right axis on Figure \ref{fig:scattersimu}. It permits the evaluation of the difference between the simulated and reconstructed field which is seen to be smaller than the variance of the simulated field itself in all but the catalogs of clusters cases.\\

\begin{figure*}
\vspace{-1.4cm}
\includegraphics[width=0.9 \textwidth]{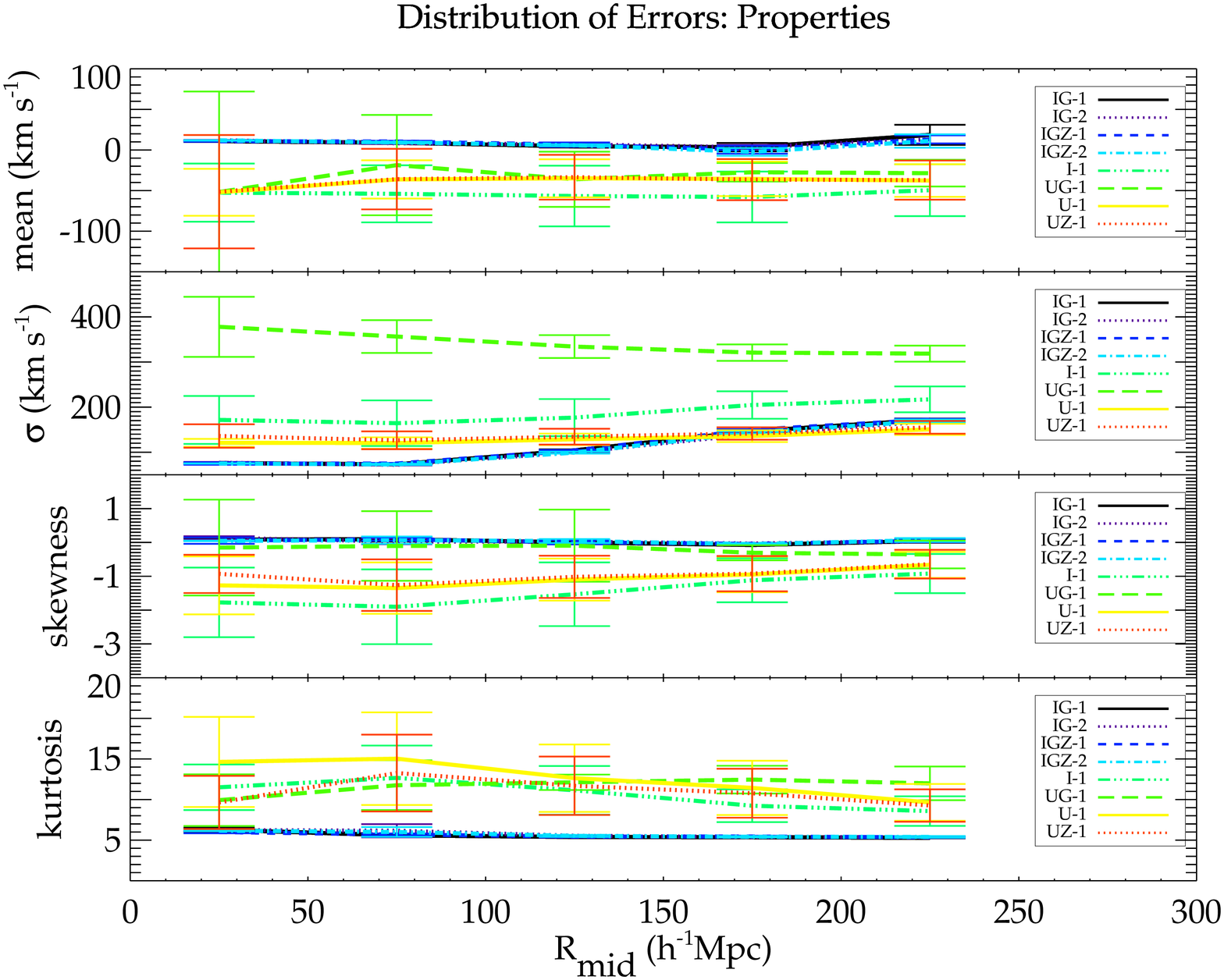}
\caption{Properties of the distribution of residuals between simulated and reconstructed fields. Namely the average properties of the distribution of errors in the reconstructions are shown. From top to bottom, averages and standard deviations of the mean, standard deviation, skewness and kurtosis of the residual distributions in shells of `middle' radius R$_{mid}$, i.e. the residuals in cells at a distance d such that R$_{mid}$-25$\le$~d~$<$~R$_{mid}$+25~\hMpc\ are considered. Each colored linestyle corresponds to one mock type, whose short names are given in Table \ref{Tbl:1}.}
\label{fig:distrierr}
\end{figure*}

Residuals between simulated and reconstructed fields can be studied in more detail. Figure \ref{fig:distrierr} shows the mean properties (and their standard deviation) of the residual distributions in the same shells as described above. From top to bottom, the averages and standard deviations of the mean, standard deviation, skewness and kurtosis. Note that skewness and kurtosis are employed here and in the rest of the paper for third and fourth standardized moments, more precisely kurtosis is used for 'excess kurtosis'. of the residual distributions per mock type are plotted. Unconditionally, inhomogeneous grouped mocks result in quasi-Gaussian residual distributions (skewness and kurtosis are almost null: about 0.1 or less in absolute value and about 1 or less respectively) with the smallest standard deviation (about 80 to 170~\kms) and a mean close to zero (less than 10~\kms\ in absolute value), i.e. reconstructions are not biased towards an infall or outflow onto the local Volume. On the contrary, all the other mock types (homogeneous or ungrouped) present on average a distribution of residuals with a negative mean of about 50~\kms\ implying that Wiener Filter reconstructed velocities are biased towards larger values with respect to the velocities of the reference simulation.  Although catalogs of clusters result in residual distributions with the largest standard deviation on average (more than 300~\kms), these distribution do not appear to be asymmetric in a quantitative way (skewness about 0.3 in absolute value). However, they are quite flat (kurtosis about 5). Finally, homogeneous and inhomogeneous catalogs that are ungrouped lead to residual distributions that, in addition to being quite flat, are asymetric (skewness about -1, -2). These observations hold for all the shells considered here. To summarize, inhomogeneous grouped catalogs result in non-biased reconstructions with on average an almost Gaussian distribution of errors presenting the smallest scatter.\\

\begin{figure*}
\vspace{-2.4cm}
\includegraphics[width=0.9 \textwidth]{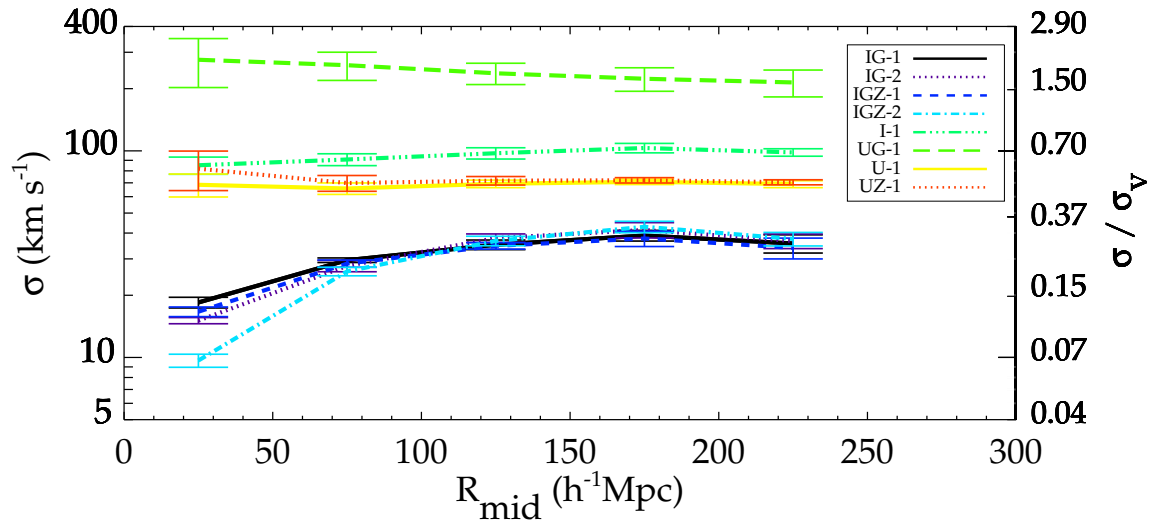}\\ 

\vspace{-1.8cm}
\caption{Means (lines) and standard deviations (error bars) of 1-$\sigma$ scatters obtained comparing between themselves reconstructed velocity fields of a same given mock category as a function of the `middle' radius of the compared shells, i.e. cells at a distance d such that R$_{mid}$-25$\le$~d~$<$~R$_{mid}$+25~\hMpc\ are compared. Each colored linestyle corresponds to one mock type, whose short names are given in Table \ref{Tbl:1}.}
\label{fig:scatterwf}
\end{figure*}

Next, the reconstructions obtained in one category are compared between themselves to estimate directly the variance due to the selection of data points. We repeat the same cell-to-cell analysis but now WF reconstructions are compared two by two instead of a comparison between a reconstructed field and the field of the reference simulation. The means (lines) and standard deviations (error bars) of the 1-$\sigma$ scatters are shown in Figure \ref{fig:scatterwf} as a function of R$_{mid}$ of the shells within which the WF reconstructions are compared. Each color represents one type of mocks as given by the legend and the meaning of the short names are explained in Table \ref{Tbl:1}.

Again, the smallest 1-$\sigma$ scatters are obtained for the non-uniformly distributed mocks (about 35~$\pm$~4~\kms\ for the largest shell down to 15~$\pm$~1~\kms\ for the inner shell) and the largest ones are obtained for the mocks of clusters (more than 200~$\pm$~30~\kms). These observations are again in agreement with expectations as the clusters do not feel their own gravitational field inducing the actual flows in the field. Thus from one selection of clusters to another, a large variation in the reconstructed field is expected. While again the variance between the reconstructions from mocks with and without a ZOA is not drastically different, there is a clear importance of eliminating virial motions when a large number of data is available in a region: larger scatters are measured for the catalogs mimicking the ungrouped version of {\it cosmicflows-2} than for those mimicking the grouped version. There are benefits in having a higher density of data in the center of the box rather than a uniform distribution only when the catalogs are grouped. Note that the non-uniformly distributed catalogs present reconstructions in better agreement with each other than those obtained with homogeneous datasets also when considering the largest shells which confirms the large scale correlation of the velocities. Namely if there is less variance in the center of the box, there is also less variance in the outer part of the box. We reiterate that this is true only if the catalog is released of non-linear motions (i.e. this observation is not true for the reconstructions obtained with the ungrouped inhomogeneous mocks). 

That the mean 1-$\sigma$ scatter of the inner shells for the inhomogeneous catalogs with a ZOA and twice the number of points is smaller than  for the other inhomogeneous catalogs is only an artifact due to the impossibility to significantly vary the points in the center of the box.  The ratio of the mean 1-$\sigma$ scatters to the variance of the simulated field is again added as a right axis on Figure \ref{fig:scattersimu}. The same conclusion as before can be reached: only the catalogs of clusters present ratios larger than 1. \\

To summarize, catalogs reproducing the grouped version of {\it cosmicflows-2} in most aspects are those resulting in reconstructions presenting the most neutral and Gaussian distribution of errors with respect to the reference simulation and the smallest scatters not only with the reference simulation but also between themselves even at large radii, i.e. where the randomness of the datapoint selection is higher and where there are less data points than in uniformly distributed catalogs. This is in agreement with the fact that peculiar velocities are correlated on large scales. Then as long as the center is well constrained, a good accuracy at farther distances can be reached.

To go further into the comparison, in the next subsection, we propose to establish the accuracy of the moments of the velocity fields obtained with the WF technique.

\subsection{Moments of the velocity fields}

Monopole and dipole components of the velocity fields are defined with the Taylor expansion of the velocity field $v$ at first order. The latter can be written as follows:
\begin{equation}
v_\alpha(\mathbf{r})=v_\alpha(\mathbf{r_0})+ \frac{\partial v_\alpha}{\partial r_\beta}|_{r_0}dr_\beta
\label{eqt}
\end{equation}
where $\alpha$ is $x$, $y$ or $z$. The trace of the deformation tensor $\frac{\partial v_\alpha}{\partial r_\beta}$ in equation~\ref{eqt} is the monopole term: $Tr(\Sigma)=-\frac{\nabla \vec v}{H(z)}$ \citep[e.g.][]{2014MNRAS.441.1974L}. Considering the full three dimensional (3D) velocity field, evaluated on a Cartesian grid, the bulk velocity vector is the volume weighted mean velocity field within a top-hat sphere of radius $R$ to denote a sharp cutoff. The bulk velocity is defined for instance by \citet{2015MNRAS.449.4494H} as:
\begin{equation}
V^{WF}_{bulk}(R)=\frac{3}{4\pi R^3}\int_{r<R}v^{WF}(\mathbf{r})d^3\mathbf{r}
\end{equation}

\begin{figure*}

\vspace{-1.5cm}
\includegraphics[width=0.9 \textwidth]{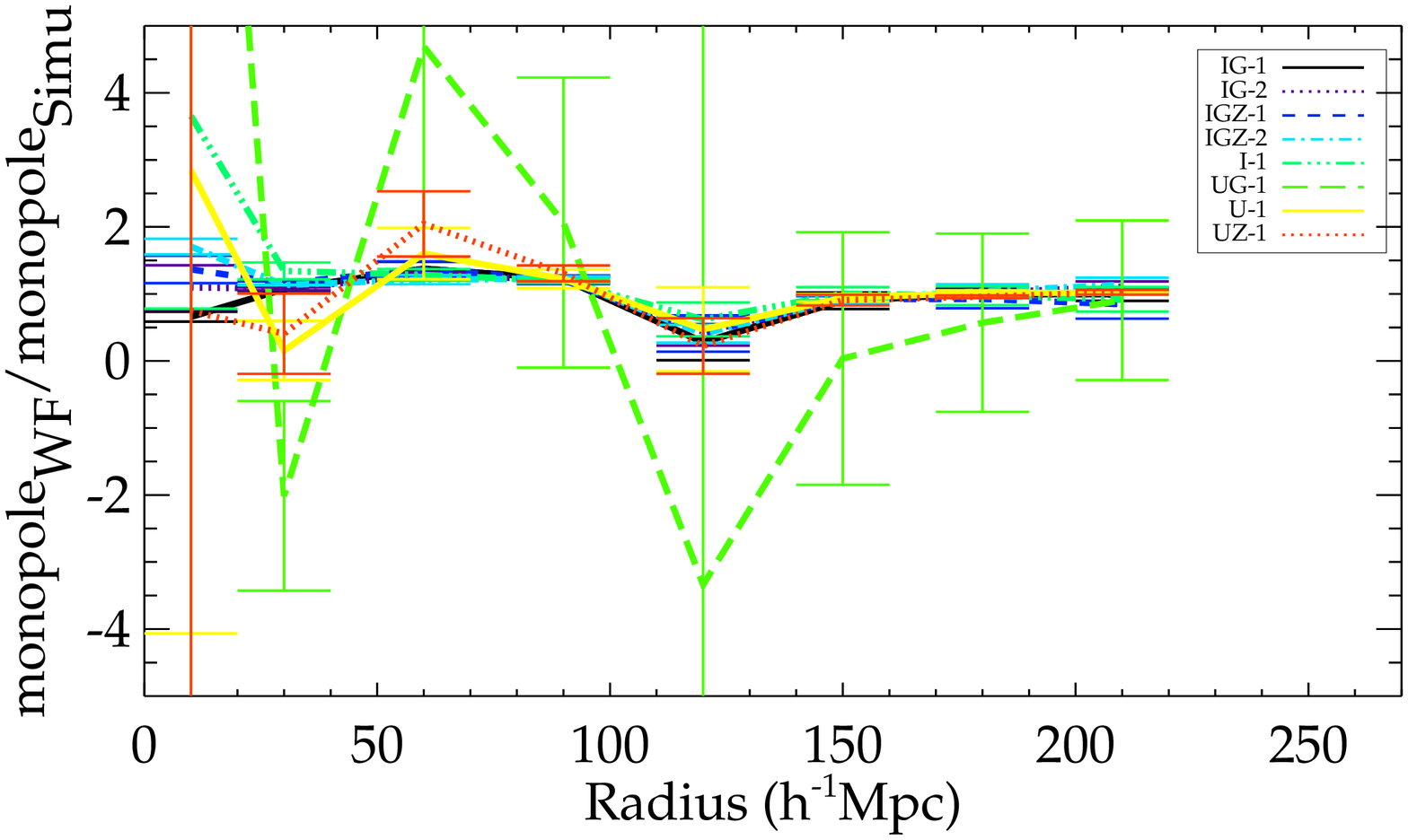}

\vspace{-0.8cm}
\includegraphics[width=0.9 \textwidth]{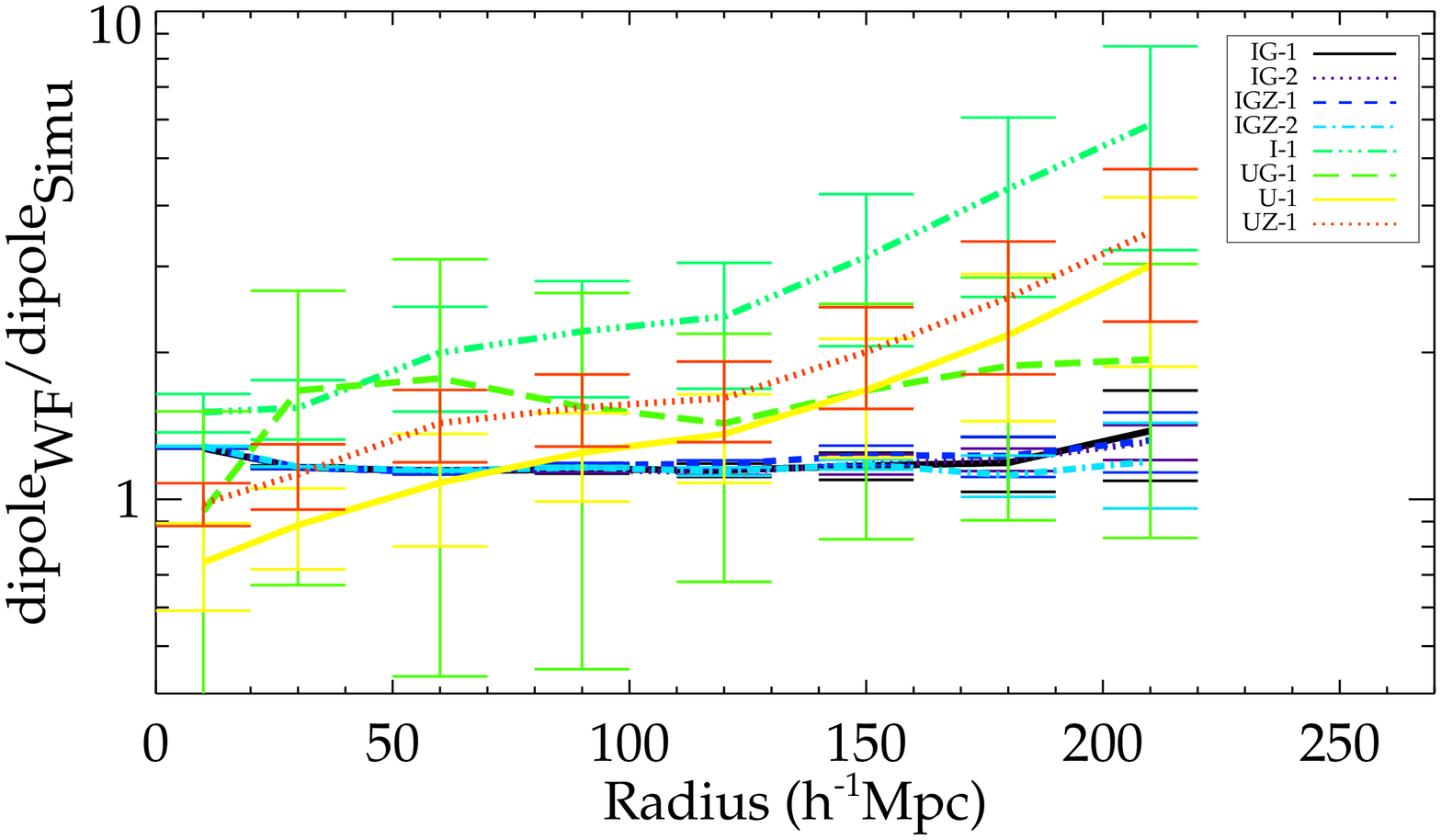}

\vspace{-0.8cm}
\includegraphics[width=0.9 \textwidth]{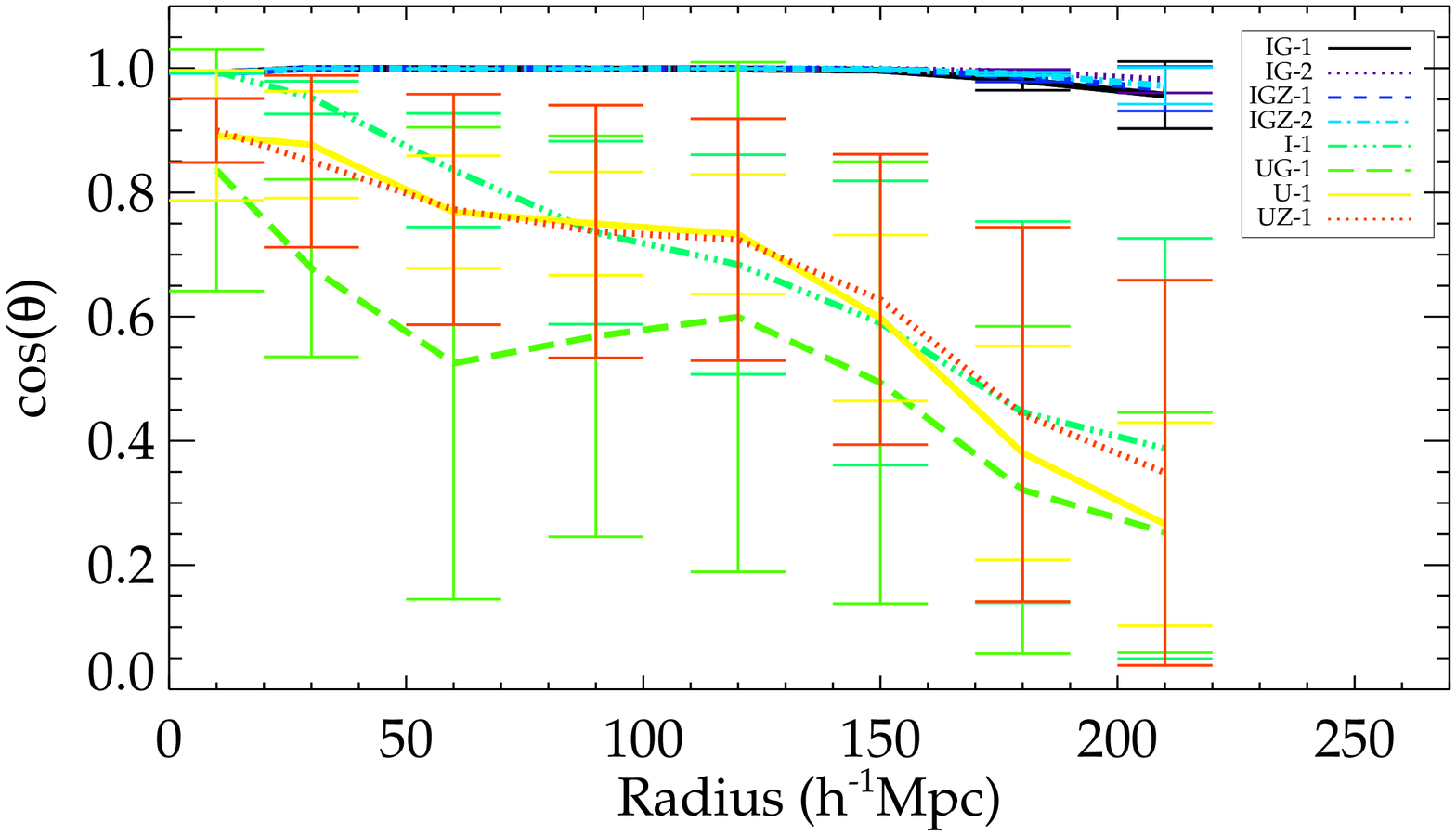}
\vspace{-0.2cm}

\caption{Mean (lines) and standard deviation (error bars) of the monopole (top) and dipole (middle) terms of the reconstructed velocity fields for each mock category divided by the monopole and dipole values of the field of the reference simulation as a function of the distance from the center of the box (observer's location). Means (lines) and standard deviations (error bars) of  the dot product or alignment (bottom) between the simulated and reconstructed dipoles as a function of the distance from the center of the box. The identification of mock categories with the letters is given in Table \ref{Tbl:1}.}
\label{fig:monodipole}
\end{figure*}

We derive the monopole and dipole components of the reference simulation and of the WF reconstructed velocity fields at different radii. Means (lines) and standard deviations (error bars) of monopole (top) and dipole (middle) values of reconstructed fields divided by those of the reference simulation as a function of the radius (distance from the center of the box) are derived and reported on Figure \ref{fig:monodipole} for each mock category (one color and linestyle per category). Mocks with non-uniform distributions have velocity moments closer to the reference simulation in agreement with the fact that their velocity fields are the best match to the reference velocity field. For the dipole term, this is not applicable to ungrouped mocks because of the non-linear motions.

More precisely, the non-uniformly distributed mocks give reconstructions with monopole terms consistent with those of the simulation: Their proportionality factor is between 0.5 and 2.0 on a large range of radii.  Regarding the uniformly distributed mocks, they have monopole terms that can vary grandly around that of the reference simulations ; especially in the case of the catalogs of clusters, where not only the monopole terms between the reconstructions obtained with the different mocks differ completely (standard deviations as high as 10), but also the ratio of the simulated to reconstructed monopole terms can reach values as high as 16. Note that on the Figure, the appropriate range chosen for visibility does not extend to 16.  

Regarding the dipole terms, the ratio between values found with the reconstructions and those of the simulation are between 1.0 and 1.3 out to 200~\hMpc\ for the grouped non-uniformly distributed mocks and present very small variances (less than 0.3). Thus the WF reconstruction in that case gives a correct bulk flow up to 200~\hMpc\ in agreement with the claims of \citet{2015MNRAS.449.4494H}. On the other hand, the ungrouped inhomogeneous mocks give values too large already at small radii (factor 2 at 60~\hMpc\ up to a factor 6 at 200~\hMpc). As for the homogeneous mocks, they start by giving reconstructions with reasonable dipole values (for radii up to 30~\hMpc) but quickly reach values 2-3 times too large with variances (about 0.4-0.5) larger than for the reconstructions obtained with the inhomogeneous mocks. 

\begin{figure*}
\vspace{-3cm}
\includegraphics[width=0.9 \textwidth]{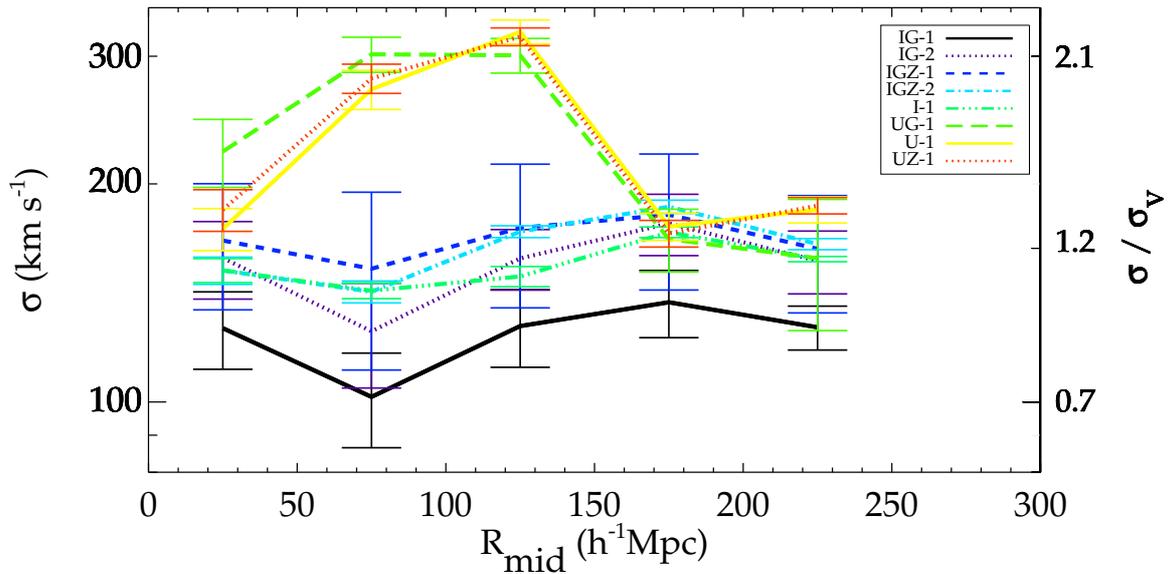}
\vspace{-2cm}

\caption{Means (lines) and standard deviations (error bars) of 1-$\sigma$ scatters obtained comparing the velocity field of the reference simulation and those obtained with the WF technique applied to the different mocks \emph{with errors} as a function of the `middle' radius of the compared shells, i.e. grid cells at a distance d such that R$_{mid}$-25$\le$~d~$<$~R$_{mid}$+25~\hMpc\ are compared. Each colored linestyle corresponds to one type of mocks. Short names of Mock types are given in Table \ref{Tbl:1}.}
\label{fig:scattersimuerr}
\end{figure*}

To investigate further the accuracy of the reconstructed dipole, we look at the alignment between the simulated and the reconstructed vectors in the bottom panel of Figure \ref{fig:monodipole}. They are extremely well aligned when considering all the inhomogeneous grouped catalogs up to distances of 200~\hMpc\ with an almost inexistent scatter. In contrast, the vectors are largely misaligned with large discrepancies and scatters for the other mocks. Considering the catalogs of clusters, the angle between the simulated and reconstructed vectors probes a large range of values and the two vectors are clearly never aligned even in the center of the box.  This clear misalignment implies that clusters are not good tracers of the large scale velocity dipole. It permits to see in a completely new way the misalignment between the CMB and the vector inferred by clusters highlighted by \citet{1994ApJ...425..418L}. In view of the results of this paper, field galaxies could reconcile the discrepancy between the dipoles' direction.

\subsection{Brief excursion in the uncertainty regime}
 A detailed study of the impact of uncertainties and their distribution in radial peculiar velocity catalogs used to reconstruct the local Universe is beyond the scope of this paper. Nevertheless, it is interesting to check to which extent our findings hold for observational catalogs with errors. Although any observer aims at providing error-free catalogs, the latter are always affected by statistical and systematic errors. To simplify the problem, we disregard the systematic errors. In light of ongoing and future \emph{all sky} surveys (e.g ASKAP, SKA, WISE, Pan-STARRS) this assumption is relatively realistic: the risks of direction-dependent errors for a given catalog will not be problematic anymore. Although statistical errors will decrease dramatically, they will still be present and cannot be strictly disregarded. However, their distribution in a given catalog of data can be quite complex. They depend on multiple factors such as the distribution of object distances to the observers, the type of observations or more precisely of distance indicators used as well as the number of distance measurements for a given object. 
 
To be realistic without adding an extra layer of complexity, we settle for a 15\% scatter Gaussian distribution of fractional errors on distances regardless of the mock type considered. The 15\% value choice is based on the fact that errors in the benchmark catalog ranged mostly between $\sim$10 and 20\%. This distribution of errors, that are proportional to the distances, allow us to take into account the fact that usually observations of objects further away are more uncertain than those of nearby objects. However, the 15\% scatter imposed whatever the mock type removes the benefit of having more data points or more observations per object that would reduce the global uncertainties. In addition, it does not take into account the fact that inhomogeneous catalogs have a smaller mean distance than homogeneous catalogs which in principle could lead to a smaller scatter of the fractional error distribution (more accurate distance indicators usable nearby). In light of these observations, results obtained with mocks where errors have been added give a hint at whether or not the finding hold but any result should be considered carefully before drawing firm conclusions.

Subsequently, we add errors to distances (thus to derived radial peculiar velocities) in every mock following the prescription we described above. Namely, for each mock, an array of random values distributed on a 0.15-$\sigma$ Gaussian is built. Distances with errors are distances without errors multiplied by the random values plus one. Peculiar velocities with errors are derived with these new uncertain distances. Note that we choose to use a different random array for every mock rather than fixing the random array for a given mock type. Consequently, the variance is increased between the mocks of a same type. Regardless, it might seem unfair to fix the random array only for a given mock type, we would need to fix it for the whole set of mocks. However, such a choice would aggravate the effects described above: there is a priori no reason for catalogs with a larger mean distance to have the same $\sigma$ (and a fortiori the exact same) Gaussian distribution of fractional errors as catalogs with a smaller mean distances.

WF reconstructions are obtained with the 40 mocks with errors and compared to the reference simulation as before. The average 1-$\sigma$ scatters and their standard deviations are plotted on Figure \ref{fig:scattersimuerr}. Results are as follows:\\
\begin{itemize}
\item The first completely expected observation is the increase of the 1-$\sigma$ scatters with respect to the case without errors: they are all above 100~\kms. The second observation is the decoupling of the 1-$\sigma$ scatters obtained with the mocks that have twice as many data points as \textit{cosmicflows-2} to those with only once the number of points. The trend is contrary to our expectation since the 1-$\sigma$ scatters are slightly larger in the former case (about 150~\kms) than in the latter case (about 130~\kms). This highlights typically one result that has to be considered carefully. Indeed the fact that the number of measurements decreases the global error has not been taken into consideration here.
\item Inhomogeneous catalogs with a ZOA reveal themselves to result in reconstructions with slightly higher 1-$\sigma$ scatters (about 170~\kms) than those without, reinforcing the weak impact of the ZOA on the reconstructions. This is further motivation to recent efforts made to observe close to the ZOA \citep{2012AJ....144..133S,2014MNRAS.444..527S,2014ApJ...792..129N} or even in the ZOA \citep{1994ASPC...67...99K,2006MNRAS.369.1741D,2014MNRAS.443...41W,2016AJ....151...52S,2016MNRAS.462.3386S,2016MNRAS.457.2366S,2016MNRAS.460..923R}. 
\item The 1-$\sigma$ scatters obtained with the ungrouped without ZOA inhomogeneous catalogs are higher (about 150~\kms) than those obtained with the inhomogeneous grouped mocks with once the number of points and without ZOA (about 130~\kms). This is in agreement with previous conclusions: Removing non-linear motions is essential. Note that such a result is not trivial since we applied the same 1-$\sigma$ Gaussian distribution of errors for both mock types whereas by definition the global error should have been smaller after grouping. 
\item This additionally explains why the 1-$\sigma$ scatters are similar for both ungrouped without ZOA inhomogeneous catalogs and grouped with ZOA inhomogeneous catalogs. Should we have considered a larger $\sigma$ for the distribution of errors in the ungrouped cases, the 1-$\sigma$ scatters would have been higher in agreement with previous observations without errors. 
\item Finally, the 1-$\sigma$ scatters are the largest (up to 300~\kms) for reconstructions obtained with the uniformly distributed mocks except at large radii. The 1-$\sigma$ scatters could have been smaller considering that clusters might benefit from multiple measurements, hence smaller errors. However, such a statement has to be counterbalanced by the fact that nearby measurements benefit in general from higher accuracy due to more precise distance indicators. In addition, even if the 1-$\sigma$ scatter could be slightly decreased, they are twice as large as those obtained with other mocks. 
\end{itemize}

To summarize, our previous conclusions based on mocks without error stand: the inhomogeneous grouped mock catalogs constitute an excellent sampling for the Wiener Filter technique.


\section{Conclusion}

The Wiener Filter (WF) permits reconstructing velocity fields with catalogs of radial peculiar velocities.  These fields are pathways to study the local Large Scale Structure and its dynamics. Thus, identifying the survey sampling that leads to the best reconstruction in the WF framework is essential. Applying the WF technique to different mock catalogs built out of a reference simulation, we are able to determine the optimal sampling of the observational dataset to recover the best quality reconstruction with the WF technique. By extension, it offers the possibility to design strategically the observational plan to measure peculiar velocities and thus build surveys.

Leading comparisons, including residual distributions, cell-to-cell and bulk velocities, between WF reconstructions obtained with catalogs of clusters, grouped catalogs (these latter differing from the former by the presence of field galaxies while galaxies in high density regions are collapsed into one point), non-uniformly and homogeneously distributed datasets drawn from a constrained simulation and that very same simulation, we find that the best quality reconstruction with a fixed number of data points is obtained with catalogs that present the following properties:
\begin{itemize}
\item they are grouped, i.e. non-linear motions are suppressed. It is an expected result as the WF is a linear method: degrading the information contained in the dataset is essential to preserve the quality of the reconstruction, 
\item they also contain field galaxies to measure flows due to the gravitational field induced by large structures like clusters, 
\item they are non-uniformly distributed, in particular they present a decreasing number of data points with the distance from the observer.
\end{itemize}

Note that there is no particular advantage at a constant number of points for a survey to be uniformly distributed especially when the object of study is the bulk flow. As a matter of fact, non-uniformly distributed datasets, as described above, result in as good reconstructions as homogeneously distributed catalogs provided that galaxies have been excellently grouped in clusters when appropriate. For instance, a catalog where 50\% of the data are within 60~\hMpc\ and 98\% within 150~\hMpc\ allows the reconstruction of the velocity field overall up to 250~\hMpc\ as well as a uniformly distributed catalog does over the same range of distances.  The velocity field reconstruction is even better in the inner part of the box when applying the WF to the former rather than the latter and the bulk flow can be studied with greater accuracy. 

Quantitatively, the WF technique when applied to a non-uniformly distributed grouped catalog of radial peculiar velocities without error, is able to provide a velocity field in agreement with the reference field (from which the data is drawn) at better than 100~\kms\ up to distances equal to three times the mean distance of the data. The dipole (including its direction) and monopole terms of the velocity field are in excellent agreement with that of the reference field (less than 2\% difference) up to 200~\hMpc. This is expected due to the large scale correlation of the velocities. The worst agreement is found for the catalogs of clusters, suggesting that such catalogs are not good probes of the large scale velocity field within the WF framework. In particular, they result in a complete misalignment between the simulated and the reconstructed dipole vectors. In addition, the random selection of points is not important provided that the sample covers the maximum of regions. Namely, for a given type of catalog, from a random selection of points to the other the resulting reconstructions do not vary much from one another.

To dedicate our attention entirely to the sampling issues, neither statistical nor systematic errors were added to catalogs in a first-pass. However, observational catalogs are not exempt of biases and errors. With a simple but realistic model of fractional errors on distances and a careful consideration of the results in light of the chosen error model, we show that our conclusions are unchanged when adding errors to the mocks. A more sophisticated error model applied to the optimal catalog as described above, typically with a density of measurements declining from the center to the outer part and with properly grouped cluster galaxies, demonstrates that the WF, applied to the catalogs where the biases have been minimized \citep[see e.g.][for a detailed study]{2015MNRAS.450.2644S}, approximates the underlying velocity field at 100-150~\kms\ in agreement with the findings obtained with the error model used here.

To summarize, this analysis lays the basis to design strategically future observational surveys and to build catalogs. Such surveys will lead to exquisite quality WF reconstruction of the Large Scale Structure that in turn will lead to optimal study of the local dynamics.

\section*{Acknowledgements}
We thank the anonymous referee for useful comments that helped improve substantially the manuscript. We thank Peter Creasey for a careful reading of the manuscript.
The authors gratefully acknowledge the Gauss Centre for Supercomputing e.V. (www.gauss-centre.eu) for providing
computing time on the GCS Supercomputer Jureca at JSC Juelich. JS acknowledges support from the Alexander Von Humboldt Foundation and from the Astronomy ESFRI and Research Infrastructure Cluster ASTERICS project, funded by the European Commission under the Horizon 2020 Programme (GA 653477). YH has been supported by the Israel Science Foundation (1013/12).

\appendix\newpage\markboth{Appendix}{Appendix}
\renewcommand{\thesection}{\Alph{section}}
\numberwithin{equation}{section}

\begin{appendix}
\section{The Wiener-Filter}
The Wiener Filter technique is the optimal minimal variance estimator given a dataset and an assumed prior power spectrum  \citep[and references therein]{1995ApJ...449..446Z}. The overdensity $\delta^{WF}$ and velocity $\textbf{v}^{WF}$ fields of the Wiener Filter are expressed in terms of the following correlation matrixes. For a list of M constraints $c_i$:

\begin{equation}
\delta^{WF}(\textbf{r})=\sum_{i=1}^M \langle\delta(\textbf{r})c_i\rangle \sum_{j=1}^M\langle c_i c_j+\epsilon_i^2\delta_{ij} \rangle^{-1}(c_j+\epsilon_j)
\label{eq1}
\end{equation}

\begin{equation}
 v_{\alpha}^{WF}=\sum_{i=1}^M \langle v_{\alpha}(\textbf{r})c_i\rangle\sum_{j=1}^M\langle c_i c_j+\epsilon_i^2\delta_{ij} \rangle^{-1}(c_j+\epsilon_j)\quad \mathrm{with} \quad \alpha=x,y,z
 \label{eq2}
 \end{equation}

where $c_i+\epsilon_i$ are mock or observational constraints plus their uncertainties and errors are assumed to be statistically independent. Schematically, the signal is smoothed by a factor inversely proportional to the errors. The constraints can be either densities or velocities. $\langle AB \rangle$ notations stand for the correlation functions involving the assumed prior power spectrum.\\
 
The associated correlation functions are given by: 
\[ \langle \delta(\textbf{r}\, ') v_{\alpha} (\textbf{r} \,'+\textbf{r}) \rangle  = \frac{\dot a f}{(2 \Pi)^3}\int_0^\infty \frac{ik_{\alpha}}{k^2}P(\textbf{k}) e^{-i\textbf{k}.\textbf{r}}d \textbf{k} \]
\begin{equation} \; = -\dot a f r_{\alpha} \zeta (r) \end{equation}

\[  \langle v_{\alpha}(\textbf{r} \,')v_{\beta}(\textbf{r}\, '+\textbf{r})\rangle \;\; = \frac{(\dot a f)^2}{(2\Pi)^3}\int_0^\infty \frac{k_{\alpha}k_{\beta}}{k^4}P(\textbf{k}) e^{-i \textbf{k} .\textbf{r}} d \textbf{k}  \]
\begin{equation} = (\dot a f)^2 \Psi_{\alpha\beta} \end{equation}

where P is the assumed prior power spectrum.\\

Since data sample a typical realization of the prior model, i.e. the power spectrum, $\frac{\chi^2}{d.o.f}$ should be close to 1 where $\chi^2=\sum_{i=1}^M\sum_{j=1}^M (c_i+\epsilon_i)\langle c_i c_j+\epsilon_i^2\delta_{ij} \rangle^{-1} (c_j+\epsilon_j)$ and d.o.f is the degree of freedom. However, data include non-linearities which are not taken into account in the model. Consequently, a non linear sigma ($\sigma_{NL}$) is required ($\langle c_i c_j \rangle+\delta_{ij}^k\epsilon_j^2+\delta_{ij}^k\sigma_{NL}^2$) to compensate for the non-linearities to drive $\frac{\chi^2}{d.o.f}$ closer to 1. Data dominate the reconstruction in regions where they are dense and accurate. In contrast when they are noisy and sparse, the reconstruction is a prediction based on the assumed prior model.
\end{appendix}


\bibliographystyle{mnras}

\bibliography{biblicomplete}
\label{lastpage}
\end{document}